# A New Standard Model of the Universe

Ø.Grøn

Oslo College, Department of engineering, Cort Adelers gt. 30, N-0254 Oslo, Norway

and

Institute of Physics, University of Oslo, Box 1048 Blindern, N-0316 Oslo, Norway

**Abstract**

Until about two years ago the dynamical properties of the evolution of the universe were assumed to be well described by the Einstein-DeSitter universe model, which is a flat universe model dominated by cold matter. However, the discovery that the cosmic expansion is accelerating made it clear that at the present time the evolution of the universe is dominated by some sort of vacuum energy with repulsive gravitation. The most simple type of vacuum energy is the Lorentz invariant vacuum energy (LIVE) which has constant energy density during the expansion of the universe. This type of vacuum may be represented mathematically by including a cosmological constant in Einstein's field equations. Hence, the flat Friedmann-Lemaître model, which is a universe model with cold matter and vacuum energy, has replaced the Einstein-DeSitter model as the standard model of the universe. In this article we give a pedagogical presentation of this model, including, among others, a new formula for the point of time when decelerated expansion changed into accelerated expansion.

## 1. Introduction

The dynamics of galaxies and clusters of galaxies has made it clear that far stronger gravitational fields are needed to explain the observed motions than those produced by visible matter.[1] At the same time it has become clear that the density of this dark matter is only about 30% of the critical density, although it is a prediction by the usual versions of the inflationary universe models that the density ought to be equal to the critical density.[2] Also the recent observations of the temperature fluctuations of the cosmic microwave radiation have shown that space is either flat or very close to flat.[3-5] The energy that fills up to the critical density must be evenly distributed in order not to affect the dynamics of the galaxies and the clusters.

Furthermore, about two years ago observations of supernovae of type Ia with high cosmic red shifts indicated that the expansion of the universe is accelerating.[6,7] This was explained as a result of repulsive gravitation due to some sort of vacuum energy. Thereby the missing energy needed to make space flat, was identified as vacuum energy. Hence, it seems that we live in a flat universe with vacuum energy having a density around 70% of the critical density and with matter having a density around 30% of the critical density.

Until the discovery of the accelerated expansion of the universe the standard model of the universe was assumed to be the Einstein-DeSitter model, which is a flat universe model dominated by cold matter. This universe model is thoroughly presented in nearly every text book on general relativity and cosmology. Now it seems that we must replace this model with a new "standard model" containing both dark matter and vacuum energy.

Recently several types of vacuum energy or so called *quintessence* energy have been discussed[8,9]. However, the most simple type of vacuum energy is the Lorentz invariant vacuum energy (LIVE), which has constant energy density during the expansion of the universe[10,11]. This type of energy can be mathematically represented by including a cosmological constant in Einstein's gravitational field equations. The flat universe model with cold dark matter and this type of vacuum energy is the Friedmann-Lemaître model, which is well known, but which is nevertheless not presented in the text books on relativity and cosmology.

In the present article I shall therefore offer a pedagogic presentation of this model, in order to provide an easily available source on this topic both for students and teachers of relativity and cosmology.

## 2. The flat Friedmann-Lemaître universe model

The Friedmann equations with cosmological constant for homogeneous and isotropic universe models with pressure free matter and flat space are

$$\frac{\ddot{a}}{a} = \frac{\Lambda}{3} - \frac{4\pi G}{3}\rho_M \tag{0.1}$$

and

$$H^2 = \frac{\dot{a}^2}{a^2} = \frac{\Lambda}{3} + \frac{8\pi G}{3}\rho_M \tag{0.2}$$

where $a$ is the expansion factor, $\Lambda$ the cosmological constant, $G$ Newton's gravitational constant, $\rho_M$ the density of cold pressure free matter, and $H$ is the Hubble parameter giving the velocity of an object at a distance $d$ due to the expansion of the universe according to Hubble's law, $v = Hd$. From (2.1) and (2.2) follows

$$2\frac{\ddot{a}}{a} + \frac{\dot{a}^2}{a^2} = \Lambda \tag{0.3}$$

Integration leads to

$$a\dot{a}^2 = \frac{\Lambda}{3}a^3 + K \tag{0.4}$$

where $K$ is a constant of integration. Since the amount of matter in a volume comoving with the cosmic expansion is constant, $\rho_M a^3 = \rho_{M0} a_0^3$, where the index 0 refers to measured values at the present time. Normalizing the expansion factor so that $a_0 = 1$ and comparing eqs.(2.2) and (2.4) then gives $K = (8\pi G/3)\rho_{M0}$. Introducing a new variable $x$ by $a^3 = x^2$ and integrating once more with the initial condition $a(0) = 0$ we obtain

$$a^3 = \frac{3K}{\Lambda}\sinh^2\left(\frac{t}{t_\Lambda}\right), \quad t_\Lambda = \frac{2}{\sqrt{3\Lambda}} \tag{0.5}$$

The vacuum energy has a constant density $\rho_\Lambda$ given by

$$\Lambda = 8\pi G \rho_\Lambda \tag{0.6}$$

The critical density, which is the density making the 3-space of the universe flat, is

$$\rho_{cr} = \frac{3H^2}{8\pi G} \tag{0.7}$$

The relative density, i.e. the density measured in units of the critical density, of the matter and the vacuum energy, are respectively

$$\Omega_M = \frac{\rho_M}{\rho_{cr}} = \frac{8\pi G \rho_M}{3H^2} \tag{0.8}$$

$$\Omega_\Lambda = \frac{\rho_\Lambda}{\rho_{cr}} = \frac{\Lambda}{3H^2} \qquad (0.9)$$

Since the present universe model has flat space, the total density is equal to the critical density, i.e. $\Omega_M + \Omega_\Lambda = 1$. In terms of the values of the relative densities at the present time the expression for the expansion factor takes the form

$$a = A^{1/3} \sinh^{2/3}\left(\frac{t}{t_\Lambda}\right), \qquad A = \frac{\Omega_{M0}}{\Omega_{\Lambda 0}} = \frac{1 - \Omega_{\Lambda 0}}{\Omega_{\Lambda 0}} \qquad (0.10)$$

Using the identity $\sinh(x/2) = \sqrt{(\cosh x - 1)/2}$ this expression may be written

$$a^3 = \frac{A}{2}\left[\cosh\left(\frac{2t}{t_\Lambda}\right) - 1\right] \qquad (0.11)$$

The age $t_0$ of the universe is found from $a(t_0) = 1$, which by use of the formula $\arctan h\, x = \arcsin h(x/\sqrt{1-x^2})$, leads to the expression

$$t_0 = t_\Lambda \arctan h\sqrt{\Omega_{\Lambda 0}} \qquad (0.12)$$

Inserting typical values $t_0 = 15 \cdot 10^9\ years$, $\Omega_{\Lambda 0} = 0.7$ we get $A = 0.43$, $t_\Lambda = 12.5 \cdot 10^9\ years$. With these values the expansion factor is $a = 0.75 \sinh^{2/3}(1.2\ t/t_0)$. This function is plotted in Fig.1.

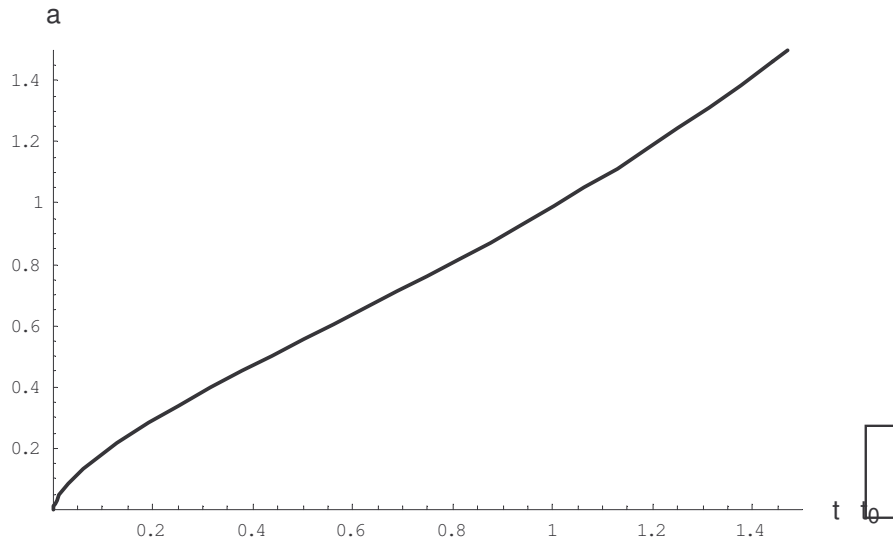

Fig.1. The expansion factor as function of cosmic time in units of the age of the universe.

The Hubble parameter as a function of time is

$$H = (2/3t_\Lambda)\coth(t/t_\Lambda) \qquad (0.13)$$

Inserting $t_0 = 1.2 t_\Lambda$ we get $H t_0 = 0.8 \coth(1.2 t/t_0)$, which is plotted in Fig.2

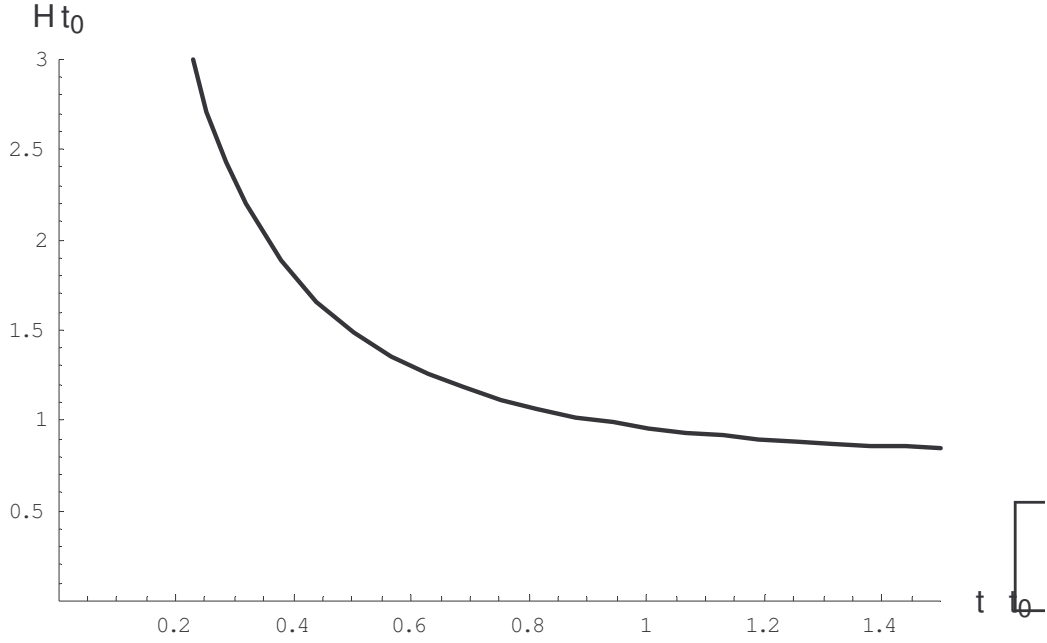

Fig.2. The Hubble parameter as function of cosmic time.

The Hubble parameter decreases all the time and approaches a constant value $H_\infty = 2/3t_\Lambda$ in the infinite future. The present value of the Hubble parameter is

$$H_0 = \frac{2}{3t_\Lambda \sqrt{\Omega_{\Lambda 0}}} \qquad (0.14)$$

The corresponding Hubble age is $t_{H0} = (3/2)t_\Lambda \sqrt{\Omega_{\Lambda 0}}$. Inserting our numerical values gives $H_0 = 64\, km/\sec Mpc^{-1}$ and $t_{H0} = 15.7 \cdot 10^9\, years$. In this universe model the age of the universe is nearly as large as the Hubble age, while in the Einstein-DeSitter model the corresponding age is $t_{0ED} = (2/3)t_{H0} = 10.5 \cdot 10^9\, years$. The reason for this difference is that in the Einstein-DeSitter model the expansion is decelerated all the time, while in the Friedmann-Lemaître model the repulsive gravitation due to the vacuum energy have made the expansion accelerate lately (see below). Hence, for a given value of the Hubble parameter the previous velocity was larger in the Einstein-DeSitter model than in the Friedmann-Lemaître model.

The ratio of the age of the universe and its Hubble age depends upon the present relative density of the vacuum energy as follows,

$$\frac{t_0}{t_{H0}} = H_0 t_0 = \frac{2}{3} \frac{\operatorname{arctan} h \sqrt{\Omega_{\Lambda 0}}}{\sqrt{\Omega_{\Lambda 0}}} \qquad (0.15)$$

This function is depicted graphically in Fig.3.

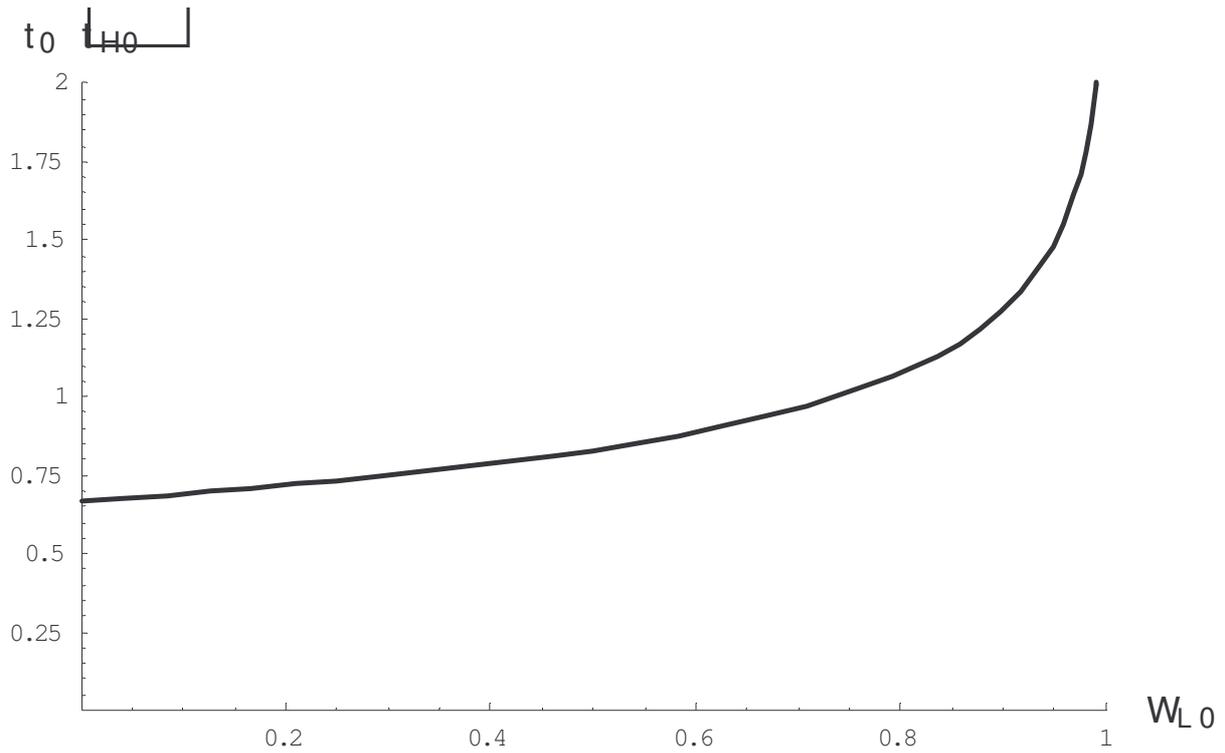

Fig.3. The ratio of the age of the universe and the Hubble age as function of the present relative density of the vacuum energy.

The age of the universe increases with increasing density of vacuum energy. In the limit that the density of the vacuum approaches the critical density, there is no dark matter, and the universe model approaches the DeSitter model with exponential expansion and no Big Bang. This model behaves in the same way as the Steady State cosmological model and is infinitely old.

The red shift of light emitted at a cosmic time $t$ and observed at $t_0$ is

$$z = \frac{a(t_0)}{a(t)} - 1 = \frac{\sinh^{2/3} 1.2}{\sinh^{2/3}(1.2 t / t_0)} - 1 \qquad (0.16)$$

Inverting this equation gives the emission time (in billion years) of light with red shift $z$ for a 15 billion years old universe

$$t = 12.5 \cdot \operatorname{arcsin} h\left(\frac{1.5}{(1+z)^{1.5}}\right) \qquad (0.17)$$

The emission time as function of red shift is shown in figure 4.

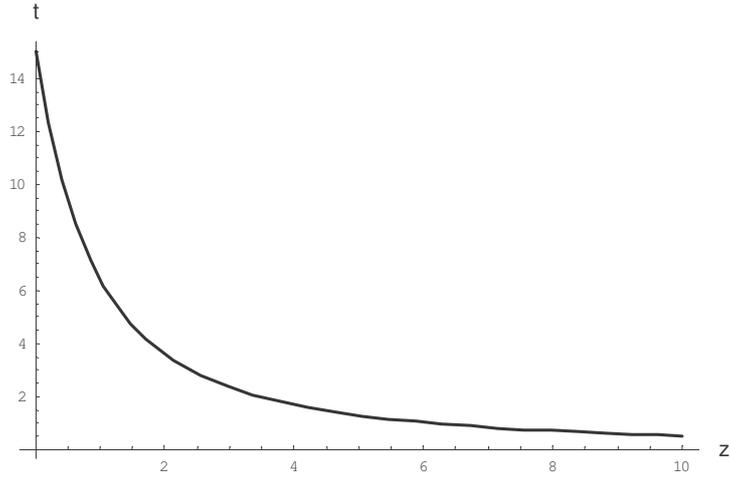

Fig.4. Emission time (in billions years) of light with red shift $z$ in a 15 billion years old universe.

A dimensioness quantity representing the rate of change of the cosmic expansion velocity is the deceleration parameter, which is defined as $q = -\ddot{a}/aH^2$. For the present universe model the deceleration parameter as a function of time is

$$q = \frac{1}{2}\left[1 - 3\tanh^2(t/t_\Lambda)\right] \qquad (0.18)$$

which is shown graphically in Fig.5.

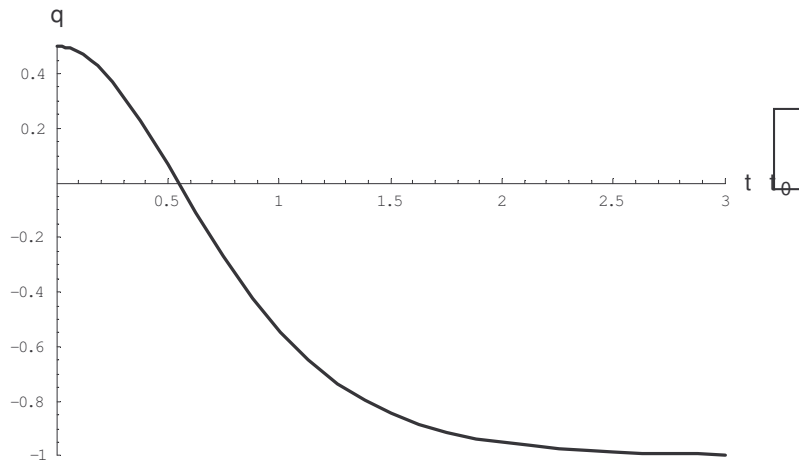

Fig.5. The deceleration parameter as function of cosmic time.

The inflection point of time $t_1$ when deceleration turned into acceleration is given by $q = 0$. This leads to

$$t_1 = t_\Lambda \operatorname{arctan} h(1/\sqrt{3}) \qquad (0.19)$$

or expressed in terms of the age of the universe

$$t_1 = \frac{\operatorname{arctan} h(1/\sqrt{3})}{\operatorname{arctan} h\sqrt{\Omega_{\Lambda 0}}} t_0 \tag{0.20}$$

The corresponding cosmic red shift is

$$z(t_1) = \frac{a_0}{a(t_1)} - 1 = \left(\frac{2\Omega_{\Lambda 0}}{1-\Omega_{\Lambda 0}}\right)^{1/3} - 1 \tag{0.21}$$

Inserting $\Omega_{\Lambda 0} = 0.7$ gives $t_1 = 0.54 t_0$ and $z(t_1) = 0.67$.

The results of analysing the observations of supernova SN 1997 at $z = 1.7$, corresponding to an emission time $t_e = 0.30 t_0 = 4.5 \cdot 10^9 \ years$, have provided evidence that the universe was decelerated at that time[12]. M.Turner and A.G.Riess[13] have recently argued that the other supernova data favour a transition from deceleration to acceleration for a red shift around $z = 0.5$.

Note that the expansion velocity given by Hubble's law, $v = Hd$, always decreases as seen from Fig.2. This is the velocity away from the Earth of the cosmic fluid at a fixed physical distance $d$ from the Earth. The quantity $\dot{a}$ on the other hand, is the velocity of a fixed fluid particle comoving with the expansion of the universe. If such a particle accelerates, the expansion of the universe is said to accelerate. While $\dot{H}$ tells how fast the expansion velocity changes at a fixed distance from the Earth, the quantity $\ddot{a}$ represents the acceleration of a free particle comoving with the expanding universe. The connection between these two quantities is $\ddot{a} = a(\dot{H} + H^2)$.

The ratio of the inflection point of time and the age of the universe, as given in eq.(2.18), is depicted graphically as function of the present relative density of vacuum energy in Fig.6.

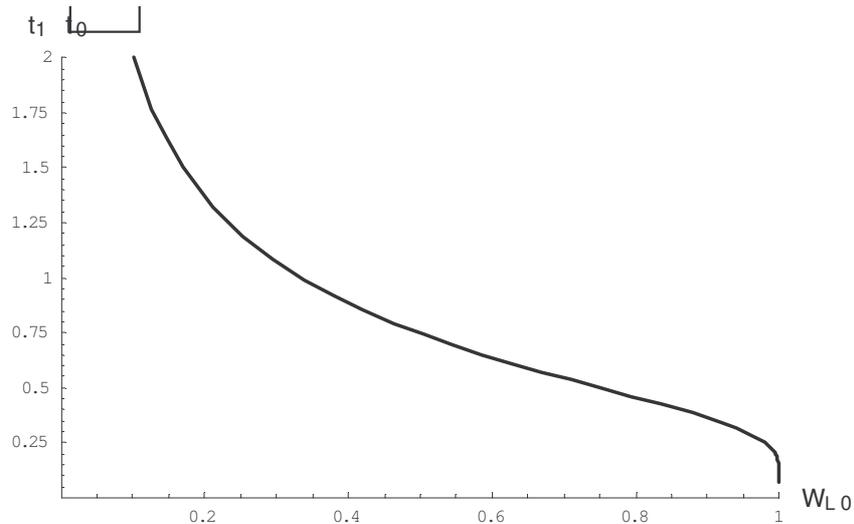

Fig.6. The ratio of the point of time when cosmic decelerations turn over to acceleration to the age of the universe.

The turnover point of time happens earlier the greater the vacuum density is. The change from deceleration to acceleration would happen at the present time if $\Omega_{\Lambda 0} = 1/3$.

The red shift of the inflection point given in eq.(2.19) as a function of vacuum energy density, is plotted in Fig.7.

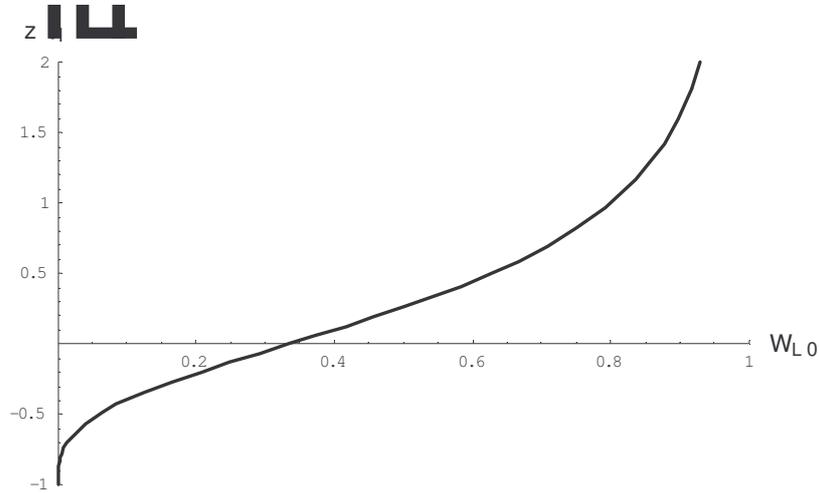

Fig.7. The cosmic red shift of light emitted at the turnover time from deceleration to acceleration as function of the present relative density of vacuum energy.

Note that the red shift of future points of time is negative, since then $a > a_0$. If $\Omega_{\Lambda 0} < 1/3$ the transition to acceleration will happen in the future.

The critical density is

$$\rho_{cr} = \rho_\Lambda \tanh^{-2}(t/t_\Lambda) \qquad (0.22)$$

This is plotted in Fig.8.

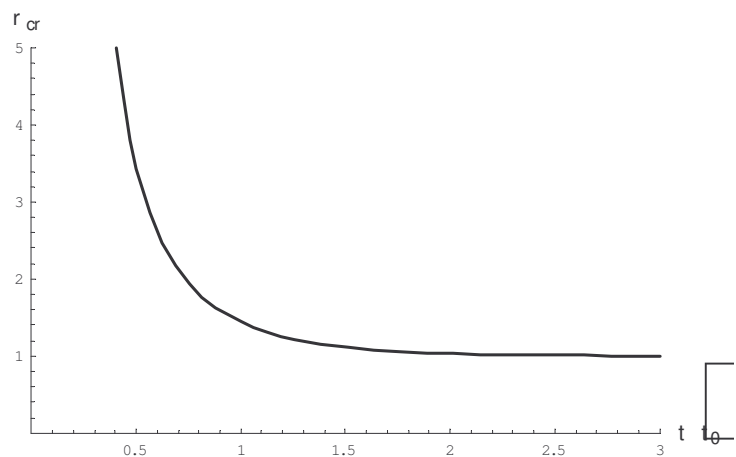

Fig.8. The critical density in units of the constant density of the vacuum energy as function of time.

The critical density decreases with time.

Eq.(2.19) shows that the relative density of the vacuum energy is

$$\Omega_\Lambda = \tanh^2(t/t_\Lambda) \tag{0.23}$$

which is plotted in Fig.9.

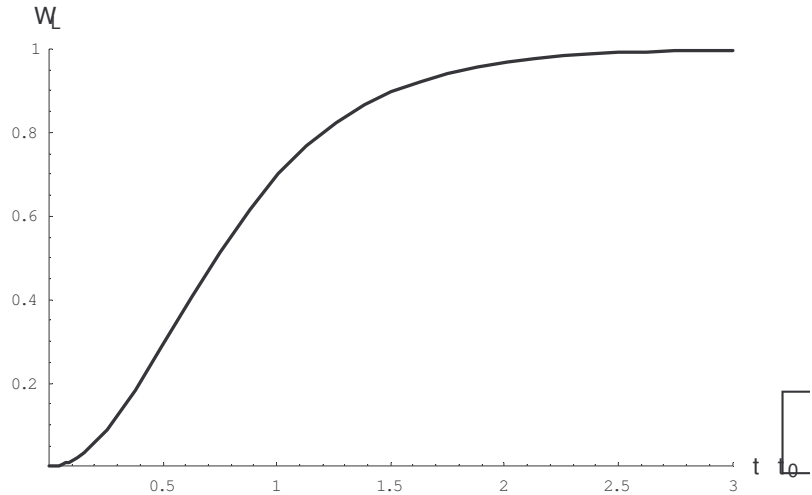

Fig.9. The relative density of the vacuum energy density as function of time.

The density of the vacuum energy approaches the critical density. Since the density of the vacuum energy is constant, this is better expressed by saying that the critical density approaches the density of the vacuum energy. Furthermore, since the total energy density is equal to the critical density all the time, this also means that the density of matter decreases faster than the critical density. The density of matter as function of time is

$$\rho_M = \rho_\Lambda \sinh^{-2}(t/t_\Lambda) \tag{0.24}$$

which is shown graphically in Fig.10.

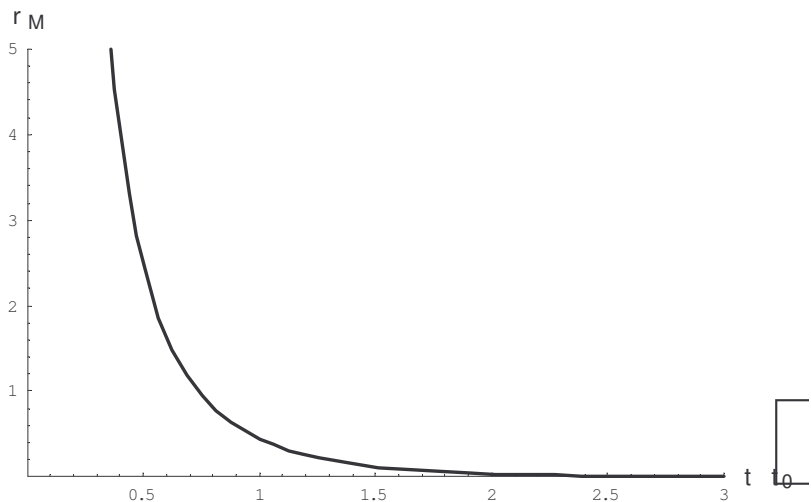

Fig.10. The density of matter in units of the density of vacuum energy as function of time.

The relative density of matter as function of time is.

$$\Omega_M = \cosh^{-2}(t/t_\Lambda) \qquad (0.25)$$

which is depicted in Fig.11.

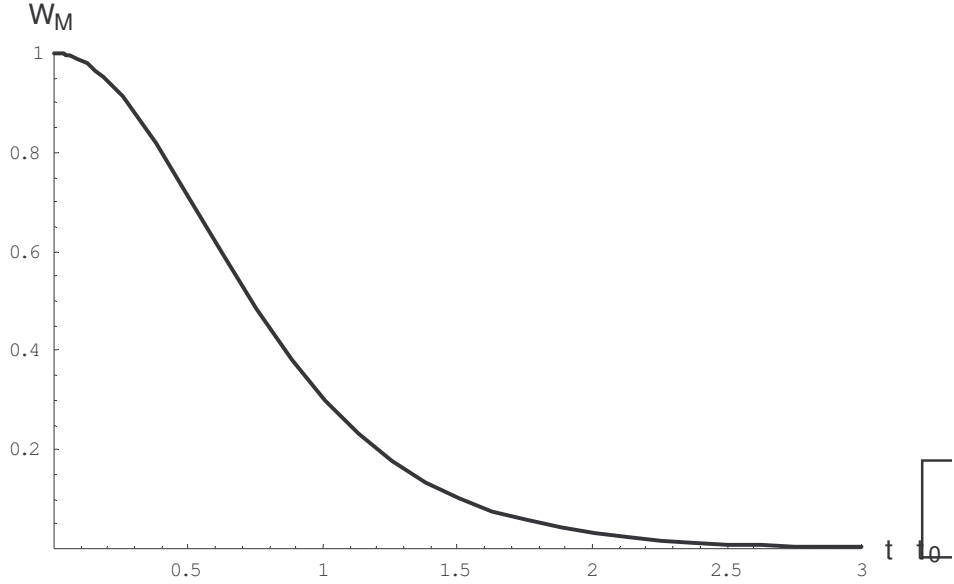

Fig.11. The relative density of matter as function of time.

Adding the relative densities of Fig.8 and Fig.10 or the expressions (2.20) and (2.22) we get the total relative density $\Omega_{TOT} = \Omega_M + \Omega_\Lambda = 1$.

The universe became vacuum dominated at a point of time $t_2$ when $\rho_\Lambda(t_2) = \rho_M(t_2)$. From eq.(2.22) follows that this point of time is given by $\sinh(t_2/t_\Lambda) = 1$. According to eq.(2.12) we get

$$t_2 = \frac{\operatorname{arcsin} h(1)}{\arctan h(\sqrt{\Omega_{\Lambda 0}})} t_0 \qquad (0.26)$$

From eq.(2.10) follows that the corresponding red shift is

$$z(t_2) = A^{-1/3} - 1 \qquad (0.27)$$

Inserting $\Omega_{\Lambda 0} = 0.7$ gives $t_2 = 0.73 t_0$ and $z(t_2) = 0.32$. The transition to accelerated expansion happens before the universe becomes vacuum dominated.

Note from eqs.(2.16) and (2.21) that in the case of the flat Friedmann-Lemaître universe model, the deceleration parameter may be expressed in terms of the relative density of vacuum only, $q = (1/2)(1 - 3\Omega_\Lambda)$. The supernova Ia observations have shown that the expansion is now accelerating. Hence if the universe is flat, this alone means that $\Omega_{\Lambda 0} > 1/3$.

## 3. Conclusion

As mentioned in the introduction, many different observations indicate that we live in a universe with critical density, where cold matter contributes with about 30 % of the density and vacuum energy with about 70 %. Such a universe is well described by the Friedmann-Lemaître universe model that have been presented above.

However, this model is not quite without problems in explaining the observed properties of the universe. In particular there is now much research directed at solving the so called *coincidence problem*. As we have seen, the density of the vacuum energy is constant during the expansion, while the density of the matter decreases inversely proportional to a volume comoving with the expanding matter. Yet, one observes that the density of matter and the density of the vacuum energy are of the same order of magnitude at the present time. This seems to be a strange and unexplained coincidence in the model. Also just at the present time the critical density is approaching the density of the vacuum energy. At earlier times the relative density was close to zero, and now it changes approaching the constant value 1 in the future. S. M. Carroll[14] has illustrated this aspect of the coincidence problem by plotting $\dot{\Omega}_\Lambda$ as a function of $\ln(t/t_0)$. Differentiating the expression (2.21) we get

$$\frac{t_\Lambda}{2}\frac{d\Omega_\Lambda}{dt} = \frac{\sinh(t/t_\Lambda)}{\cosh^3(t/t_\Lambda)} \qquad (0.28)$$

which is plotted in Fig.12.

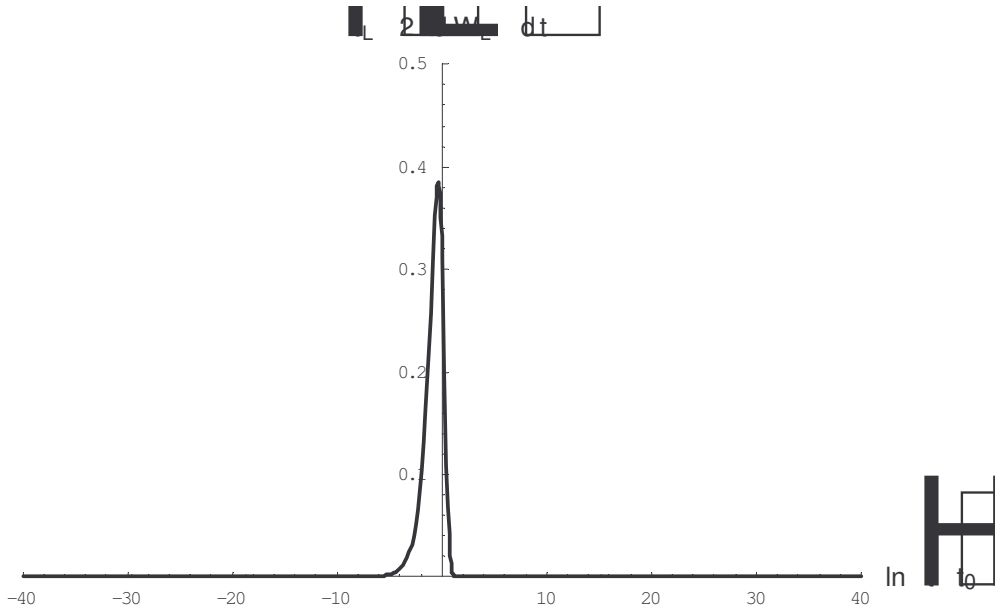

Fig.12. Rate of change of $\Omega_\Lambda$ as function of $\ln(t/t_0)$. The value $\ln(t/t_0) = -40$ corresponds to the cosmic point of time $t_0 \sim 1\sec$.

Putting $\ddot{\Omega}_\Lambda = 0$ we find that the rate of change of $\Omega_\Lambda$ was maximal at the point of time $t_1$ when the deceleration of the cosmic expansion turned into acceleration. There is now a great activity in order to try to explain these coincidences by introducing more general forms of vacuum energy called quintessence, and with a density determined dynamically by the evolution of a scalar field.[15]

However, the simplest type of vacuum energy is the LIVE. One may hope that a future theory of quantum gravity may settle the matter and let us understand the vacuum energy. In the meantime we can learn much about the dynamics of a vacuum dominated universe by studying simple and beautiful universe models such as the Friedmann-Lemaître model.